\documentclass[letter,superscriptaddress,twocolumn, pra,showkeys,showpacs]{revtex4-1}

	\usepackage{amsmath}
	\usepackage{makeidx}
	\usepackage{amsfonts}
	\usepackage[ansinew]{inputenc}
	\usepackage[usenames,dvipsnames]{pstricks}
    \usepackage{graphicx}
    \usepackage{subfigure}
	\usepackage{pst-grad} 
	\usepackage{pst-plot} 
	\usepackage{sidecap}

	\usepackage[colorlinks,hyperindex]{hyperref}
	
	\hypersetup
	{
		colorlinks,%
		citecolor=black,%
		linkcolor=black,%
		urlcolor=black,%
	}

	\newcommand{\ket}[1]{\left| #1 \right\rangle}
	\newcommand{\bra}[1]{\left\langle #1 \right|}

\makeindex

\begin{document}

\title{Entanglement Swapping Between Dissipative Systems}

\author{A. Nourmandipour}
\email{anoormandip@stu.yazd.ac.ir}
\affiliation{Atomic and Molecular Group, Faculty of Physics, Yazd University, Yazd  89195-741, Iran}
\author{M. K. Tavassoly}
\email{mktavassoly@yazd.ac.ir}
\affiliation{Atomic and Molecular Group, Faculty of Physics, Yazd University, Yazd  89195-741, Iran}
\affiliation{Photonic Research Group, Engineering Research Center, Yazd University, Yazd  89195-741, Iran}

\date{today}

\begin{abstract}
In this paper, we investigate the possibility of entanglement swapping between two distinct qubits coupled to their own (in general) non-Markovian environments. This is done via Bell state measurement performing on the photons leaving the dissipative cavities. In the continuation, we introduce the concept of entangling power to measure the average of swapped entanglement over all possible pure initial states. Then, we present our results in two strong and weak coupling regimes and discuss the role of detuning parameter in each regime on the amount of swapped entanglement. We also determine the conditions in which the maximum amount of entanglement can be swapped between two qubits. It is revealed that despite of the presence of dissipation, it is possible to create long-living stationary entanglement between two qubits.
\end{abstract}


\maketitle

\section{Introduction}
Recently, a great deal of attention has been devoted to the concept of quantum entanglement \cite{Horodecki2009QuantumEntanglement} due to its various applications such as quantum cryptography \cite{Ekert1991}, quantum teleportation \cite{Braunstein1995}, superdense coding \cite{Mattle1996}, sensitive measurements \cite{Richter2007} and quantum telecloning \cite{Muarao1999}. 
There are many implementations to produce entangled states, such as trapped ions \cite{Turchette1998}, atomic ensembles \cite{Julsgaard2001}, photon pairs \cite{Aspect1981} and superconducting qubits \cite{Izmalkov2004}. However, it is well-known that the interaction of atoms with various types of cavity field (with additional interaction terms such as Kerr medium, etc.) is an efficient source of entanglement \cite{Baghshahi2014}, a model which is called the Jaynes-Cummings model (JCM) \cite{Jaynes1963}. 
It relies on the mutual coupling between a two-level atom and a single-mode quantized field in the rotating wave approximation. 

On the other hand, it has been put forward this idea that it is possible to create entanglement between subsystems distributed over long distances without any common past. In such cases, one could think of entangling the subsystems by constructing a more general system, with help of two (or more) another entangled quantum subsystems, a phenomenon which is called quantum swapping \cite{Zukowski1993}. 
This notion has originally been proposed to  swap the entanglement between a pair of  particles \cite{Zukowski1993} and later was generalized to the multi-particle quantum systems \cite{Bose1998}. It has also been shown that it is possible to implement the quantum swapping for continuous variable systems \cite{Polkinghorne1999}. The experimental demonstration of  unconditional entanglement swapping for continuous variables has been investigated in \cite{Jia2004}. The possibility of optimization of entanglement purification via entanglement swapping has been studied in \cite{Shi2000}. Entanglement swapping in two independent JCM has been discussed in \cite{Liao2011}. In addition, by replacing the unknown state with an entangled state, entanglement swapping can be considered as a special example of quantum teleportation \cite{Lee2011,Daneshmand2016}. The basic concept concealed behind the quantum swapping is the Bell state measurement (BSM) approach. This notion can be thought as a projection operator which projects the state of fields into a Bell state and leaves qubits in an entangled state \cite{Deng2006}.

However, dissipation is ever present in real physical systems. This is due to the unavoidably interaction between those systems with their surrounding environments which usually leads to loss of entanglement stored in those systems. Therefore, a lot of attentions have been paid on the theory of open quantum systems \cite{Rafiee2016,Breuer2002}. In this regard, beside considering the Lindblad master equation which is based on the temporal evolution of the density operator of the system \cite{Breuer2002}, one can deal with the time evolution of the wave function of the system instead of density operator by solving the time-dependent Schrödinger equation. Recently, this approach has been used by us to investigate the dynamics of entanglement of two qubits in separate environments \cite{nourmandipour2015novel}, two \cite{Nourmandipour2015} and an arbitrary number of qubits \cite{Nourmandipour2016,Nourmandipour2016JOSAB} in a common environment. Altogether, it seems quite logical to investigate different aspects of quantum information processing, especially quantum entanglement swapping, in the presence of dissipation. 

In this paper, we intend to study the possibility of entanglement swapping between two independent subsystems in the presence of dissipation. To end this, we consider each subsystem as a dissipative cavity, in which there is a two-level atom in each cavity interacting with a cavity field and the cavity mode itself interacts with the surrounding environment. We model the surrounding environment as a set of continuum harmonic oscillators. This allows us to obtain the exact time evolution of wave function of each atom-environment subsystem as a function of the environment correlation time and investigate the dynamics of entanglement of each subsystem outside the Markovian regime for both weak and strong coupling regimes by paying attention to the linear entropy. Then, with the help of BSM performing on the fields leaving the cavities, we show that how the produced atom-field entanglement can be swapped between field-filed and atom-atom, resulting a final (possible) atom-atom entangled state in the presence of dissipation. We then quantify the amount of entanglement via concurrence \cite{Wootters1998}. We shall then use an entangling power measure by generalizing the expression used for unitary maps \cite{Zanardi2000} in order to see on average, how much entanglement can be swapped between two atoms. The entangling power relies on the statistical average
over the initial states which establishes an input-independent dynamics of entanglement. A concept that has already been applied in many quantum systems \cite{Ye2004,Abreu2006}.

The rest of paper is organized as follow: In Sec. \ref{sec.Model}, we introduce our modelling of dissipation of the system under consideration and obtain the explicit form of the state vector of entire system at any time $t$. In Sec. \ref{sec.Ent} we investigate the dynamics of linear entropy of each subsystem. Sec. \ref{sec.Etp} deals with the dynamical behaviour of entangling power of the atom-atom state after BSM for two types of fields Bell states. Finally, in Sec. \ref{sec.Con} we draw our conclusion.

\section{Model}\label{sec.Model}

\begin{figure}[ht]
   \centering
\includegraphics[width=0.5\textwidth]{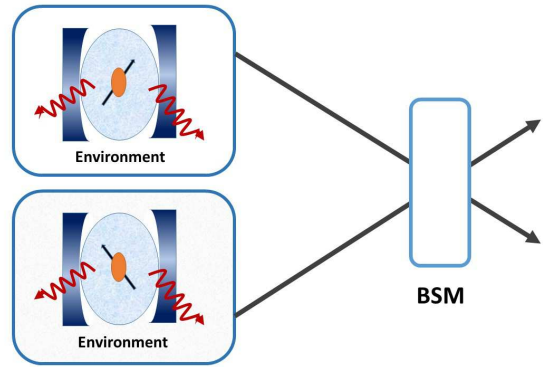}
   \caption{\label{Fig_1} Pictorial representation of the entanglement swapping. Each qubit has been placed in its own cavity in the presence of dissipation. The BSM is performed on the photons leaving the environments which leads to the establishment of entanglement between the atoms.}
  \end{figure}

The system under consideration consists of two similar but separate dissipative cavities, each contains a two-level atom with excited (ground) state $\ket{e}$ ($\ket{g}$). We model each dissipative cavity as a high-Q cavity in which the qubit interacts with a single-mode field, however, the field itself interacts with an external field which is considered as a set of continuum harmonic oscillators (see Fig. \ref{Fig_1}). The correlation between the qubit and the field in each cavity via coupling constant $g_i$ is characterized by the terms like $g_i\left( \hat{\sigma}_{i}^+\hat{a}_{i} + \hat{\sigma}_{i}^-\hat{a}_{i}^{\dagger}\right) $ in which $ \hat{\sigma}_{i}^+$ ($\hat{\sigma}_{i}^-$) is the raising (lowering) operator of the $i$th qubit and $\hat{a}_{i}$ ($\hat{a}_{i}^{\dagger}$) is the annihilation (creation) operator of the $i$th cavity field. The interaction between the cavity and the external field in the $i$th cavity can be governed by Hamiltonian 
\begin{equation}
 \begin{aligned}
 \hat{H}_{\text{I}_i}&= \omega_{c_i} \hat{a}_i^{\dagger} \hat{a}_i  \ + \ \int_0^{\infty}\! \eta \hat{B}_i^{\dagger}(\eta)\hat{B}_i(\eta) \, \mathrm{d}\eta \ \\ &+ \ \int_0^{\infty}\! \left(   G_i(\eta)\hat{a}_i^{\dagger}\hat{B}_i(\eta)\ + \ \text{H.c.} \right)  \, \mathrm{d}\eta,
  \end{aligned}
  \label{eq:int}
  \end{equation}
where $\omega_{c_i}$ is the frequency of $i$th cavity field, $G_i(\eta)$ is the coupling coefficient  which in general, is a function of frequency that connects the external world to the $i$th cavity, and $\hat{B}_i^{\dagger}(\eta)$ and $\hat{B}_i(\eta)$ are the creation and annihilation operators of the $i$th surrounding environment at mode $\eta$ which obey the commutation relation $\left[ \hat{B}_i(\eta),\hat{B}_j^{\dagger}(\eta^{'}) \right]  =\delta_{ij}\delta(\eta-\eta^{'})$. From this point of view, it can be found out  that photons in each cavity can leak out to a continuum of states, which is the source of dissipation. We shall show that this model leads to a Lorentzian spectral density for each dissipative cavity which implies the nonperfect reflectivity of the cavity mirrors. 
In the continuation, we assume that each surrounding environment possesses such a narrow bandwidth  that only a particular mode of the cavity can be excited \cite{Nourmandipour2015,Dutra2005}. This allows one to extend integrals over $\eta$ back to $-\infty$ and take $G_i(\eta)$ as a constant (equal to $\sqrt{\kappa_i/\pi}$). Then, by introducing the dressed operators $\hat{A}_i(\omega)=\alpha_i(\omega)\hat{a}_i+\int\! \beta_i(\omega,\eta)\hat{B}_i(\eta) \, \mathrm{d}\eta$ one is able to  diagonalize the Hamiltonian (\ref{eq:int}), where $\alpha_i(\omega)$ and $\beta_i(\omega,\eta)$ (in general $\in \mathbb{C}$) are obtained such that $\hat{A}_i(\omega)$ ($i=1,2$) are  annihilation operators  obeying the commutation relation  $\left[ \hat{A}_i(\omega),\hat{A}_j^{\dagger}(\omega^{'})\right]=\delta_{ij}\delta(\omega-\omega^{'})$ \cite{Fano1961,Barnett2002}. The bosonic operator $\hat{a}_i$ can be shown to be a linear combination of the dressed operators $\hat{A}_i(\omega)$ as follows \cite{Nourmandipour2015,Dutra2005}:
\begin{equation}
\hat{a}_i=\int\! \alpha_i^{*}(\omega)\hat{A}_i(\omega) \, \mathrm{d}\omega,
\label{eq:operatorA}
\end{equation}
with
\begin{equation}
\alpha_i(\omega)=\frac{\sqrt{\kappa_i/\pi}}{\omega-\omega_{c_i}+i\kappa_i}. \label{eq:alpha} 
\end{equation}
From this point of view, one can dedicate that, in each cavity the interaction between the qubit with the surrounding environment is governed by terms like $g_i\int\!\left(\hat{\sigma}_i^+\alpha_i^{*}(\omega)\hat{A}_i(\omega)+\text{H.c.}\right)\mathrm{d}\omega$.  Henceforth, the Hamiltonian describing each atom-cavity dissipative system in the rotating wave approximation and in the unit of $\hbar=1$ can be  rewritten in terms of the dressed operators as follows
\begin{equation}
\label{eq:Hfinal}
\begin{aligned}
\hat{H}_{\left( \mathrm{AF}\right)_i }&= \dfrac{1}{2}\omega_{\text{qb}_i}\hat{\sigma}_{z_i}+\int\! \omega \hat{A}_i^{\dagger}(\omega)\hat{A}_i(\omega) \, \mathrm{d}\omega \\
&+ g_i\int\!\left(\hat{\sigma}_i^+\alpha_i^{*}(\omega)\hat{A}_i(\omega)+\text{H.c.}\right) \, \mathrm{d}\omega,  
\end{aligned}
\end{equation}
in which $\omega_{\text{qb}_i}$ and $\hat{\sigma}_{z_i}$ are the atomic transition frequency and inversion population operator of the $i$th qubit, respectively. 
The time-dependent Schrödinger equation with Hamiltonian (\ref{eq:Hfinal}) can be solved  when  the environment initially is  in a vacuum state regardless of  the state of the qubit. Formally, it is convenient to work in the interaction picture. The Hamiltonian \eqref{eq:Hfinal}, in the interaction picture, is given by:
\begin{equation}
\hat{{\mathfrak V}}_{\left( \mathrm{AF}\right)_i }=e^{i\hat{H}_{\left( \mathrm{AF}\right)_i }^0t}\hat{H}_{\left( \mathrm{AF}\right)_i }^{\text{Int}}e^{-i\hat{H}_{\left( \mathrm{AF}\right)_i }^0t},
\end{equation}
in which
\begin{equation}
\begin{aligned}
\hat{H}_{\left( \mathrm{AF}\right)_i }^0&=\dfrac{1}{2}\omega_{\text{qb}_i}\hat{\sigma}_{z_i}+\int\! \omega \hat{A}_i^{\dagger}(\omega)\hat{A}_i(\omega) \, \mathrm{d}\omega, \\
\hat{H}_{\left( \mathrm{AF}\right)_i }^{\text{Int}}&=g_i\int\!\left(\hat{\sigma}_i^+\alpha_i^{*}(\omega)\hat{A}_i(\omega)+\text{H.c.}\right) \, \mathrm{d}\omega.
\end{aligned}
\label{eq:intpic}
\end{equation}
After some manipulation, the explicit form of the Hamiltonian in the interaction picture reads as:
\begin{equation}
\hat{{\mathfrak V}}_{\left( \mathrm{AF}\right)_i }=g_i\int\!\left(\hat{\sigma}_i^+\alpha_i^{*}(\omega)e^{i(\omega_{\text{qb}_i}-\omega)t}\hat{A}_i(\omega)+\text{H.c.}\right) \, \mathrm{d}\omega.
\end{equation}

It should be noted that, solving the time-dependent Schrödinger equation analytically with arbitrary initial state seems to be a very hard task, if not impossible. Without loss of generality, we assume that the two subsystems are similar, i.e., $\omega_{\text{qb}_1}=\omega_{\text{qb}_2}\equiv \omega_{\text{qb}}$, $\omega_{\text{c}_1}=\omega_{\text{c}_2}\equiv \omega_{\text{c}}$, $g_1=g_2\equiv g $ and $\kappa_1=\kappa_2\equiv \kappa$.

We assume that there is no excitation in the cavities before the occurrence of interaction and each atom is in the coherent superposition of the exited $\ket{e_i}$ and  ground state $\ket{g_i}$ as
\begin{equation}
\ket{\psi_{\mathrm{AF}}(0)}_i=\left( \cos(\theta_i/2) \ket{e_i}+\sin(\theta_i/2) e^{i\phi_i}\ket{g_i}\right)\ket{\boldsymbol{0}}_{R_i},
\label{eq:initialstate} 
\end{equation}
in which  $\ket{\boldsymbol{0}}_{R_i}=\hat{A}_i(\omega)\ket{1_{\omega}}_i$ is the multi-mode vacuum state of the $i$th environment, where $\ket{1_{\omega}}_i=\hat{A}_i^{\dagger}(\omega)\ket{\boldsymbol{0}}_{R_i}$ is the multi-mode state of the $i$th environment representing one photon at frequency $\omega$, and vacuum state in all other modes. In the above relation $\theta_i\in\left[ 0,\pi\right] $ and $\varphi_i\in\left[ 0,2\pi\right] $ for $i=1,2$.
Accordingly, the quantum state of the $i$th system at any time $t$ can be written as
\begin{equation}
\begin{aligned}
\ket{\psi_{\mathrm{AF}}(t)}_i&=C_i(t)\ket{e_i}\ket{\boldsymbol{0}}_{R_i}+D_i(t)\ket{g_i}\ket{\boldsymbol{0}}_{R_i} \\
 &+\int\! U_{\omega_i}(t)\ket{1_{\omega}}\ket{g_i} \, \mathrm{d}\omega,
\end{aligned}
\label{eq:state}
\end{equation}
where $C_i(t)$, $D_i(t)$ and $U_{\omega_i}(t)$ are unknown coefficients should be determined.
Using  time-dependent Schrödinger equation $\left( i\dot{\ket{\psi}}=\hat{{\mathfrak V}}\ket{\psi}\right) $, one arrives at the following set of coupled integro-differential equations
\begin{subequations}
\begin{eqnarray}
\dot{C}_i(t)&=&-ig\int\! \alpha^*(\omega) e^{i\delta_{\omega}t}U_{\omega_i}(t) \, \mathrm{d}\omega, \label{eq:diff1} \\
\dot{D}_i(t)&=& 0, \label{eq:diff2} \\
\dot{U}_{\omega_i}(t)&=&-ig\alpha(\omega)e^{-i\delta_{\omega}t}C_i(t), \label{eq:diff3}  
\end{eqnarray}
\end{subequations}
where $\delta_{\omega}\equiv\omega_{\text{qb}}-\omega$. The second differential equation of the above set can be easily solved as $D_i(t)=D_i(0)=\sin(\theta_i/2) e^{i\phi_i}$. After lengthy but straightforward manipulations, the following  integro-differential equation for the amplitude $C_i(t)$ may be obtained:
\begin{equation}
\label{eq:diffCi}
\dot{C}_i(t)=-\int_{0}^{t}\! f(t-t_1)C_i(t_1) \, \mathrm{d}t_1,
\end{equation}
where  $f(t-t_1)$ is the correlation function relating to the spectral density $J(\omega)$ of the environment as
\begin{equation}
f(t-t_1)=\int\! \, \mathrm{d}\omega J(\omega) e^{-i\delta_{\omega}(t-t_1)}  , \label{eq:f}
\end{equation}
in which according to Eq. (\ref{eq:alpha}) the spectral density reads as
\begin{equation}
\label{eq:spcden}
J(\omega)\equiv g^2|\alpha(\omega)|^2=\dfrac{1}{\pi}\dfrac{g^2\kappa}{ (\omega-\omega_c)^2+\kappa^2}.
\end{equation}
The parameter $\kappa$  is connected to the damping time of the environment $\tau_B\approx\kappa^{-1}$ which is much longer than its correlation time, over which the correlation functions of the reservoir vanish \cite{Gardiner2004}. On the other hand, it can be shown that the relaxation time $\tau_R$ over which the state of system changes reads as $\tau_R\approx g^{-1}$  \cite{Bellomo2007}.

A glance at  Eq. (\ref{eq:spcden}) reveals that the spectral density is a  Lorentzian distribution which implies the nonperfect reflectivity of the cavity mirrors \cite{Breuer2002}. This leads to an exponentially decaying correlation function, with $\kappa$ as the decay rate factor of the cavity as follows:
\begin{equation}\label{eq:solf}
f(t-t_1)=g^2e^{-\kappa(t-t_1)}e^{-i\Delta(t-t_1)},
\end{equation}
in which $\Delta=\omega_c-\omega_{\text{qb}}$ is the detuning parameter. We note that, by choosing special values of $\kappa$, it is possible to extract the ideal cavity and the Markovian limits. The former is obtained when $\kappa\rightarrow 0$, which leads to $J(\omega)=g^2\delta(\omega-\omega_0)$ corresponding to a constant correlation function in which $\delta(\bullet)$ is the usual Dirac delta function. In this situation, the system reduces to a $n$-qubit Jaynes-Cummings model \cite{Tavis1968} with the vacuum Rabi frequency $\Omega_R=g$. On the other hand, for small correlation times and by taking $\kappa$ much larger than any other frequency scale, the Markovian regime may be obtained. For the other generic values of $\kappa$, the model interpolates between these two limits.

Anyway, with the help of Laplace transform technique one is able to solve the integro-differential equation (\ref{eq:diffCi}) as follows:
 \begin{equation}
C_i(t)=C_i(0){\cal E}(t)\label{eq:survivalamplitude}
 \end{equation}
 in which 
 \begin{equation}\label{eq:corrfunc}
 {\cal E}(t)\equiv e^{-(i\Delta+\kappa) t/2}\left( \cosh{\left(\Omega t/ 2 \right)} +\dfrac{i\Delta+\kappa}{\Omega}\sinh{\left( \Omega t/ 2\right) }\right).
 \end{equation}
Here $\Omega=\sqrt{\kappa^2-\Omega_R^2+2i\Delta\kappa}$, in which $\Omega_R=\sqrt{\Delta^2+4g^2 }$. The obtained analytical expression for amplitude $C_i(t)$ is exact and therefore outside Markovian regime. 
It should be noticed that, the exact solution presented in \eqref{eq:survivalamplitude} for strong and weak coupling regimes is due to the Lorentzian spectral density which has been directly obtained from our modelling of dissipative cavity. For other kinds of spectral densities, only the weak coupling regime is amenable to a general analysis \cite{Kofman2000}.

\section{Entropy Evolution of the Subsystems}\label{sec.Ent}
In this section, we intend to investigate the dynamical behaviour of entanglement between qubit and its surrounding environment in each subsystems. It is well-known that the linear entropy is a promising quantity to measure the amount of entanglement between the qubit and its surrounding environment filed which is defined as \cite{Peters2004}:
\begin{equation}\label{linear}
S_{A}(\theta,\phi;t)=1-\text{Tr}\left( \hat{\rho}_{_A}^2\right) ,
\end{equation}
in which $\hat{\rho}_{_A}$ is the atomic reduced density matrix for each subsystem. The linear entropy can range between zero, corresponding to a completely pure state, and $(1-1/d)$ corresponding to a completely mixed state, in which $d$ is the dimension of the density matrix (here, $d=2$). Using Eq. (\ref{eq:state}), the explicit form of the atomic reduced density operator  at any time can be derived by tracing over environment variables results in:
\begin{equation}
\hat{\rho}_{_A}(t)=\begin{pmatrix} 
 \left| C(t)\right| ^2 & C(t)D^*(t) \\
 D(t)C^*(t) & 1-\left| C(t)\right| ^2
 \end{pmatrix},
 \label{eq:red}
\end{equation} 
in which we have dropped the subscript $i$ from coefficients $C_i(t)$ and $D_{i}(t)$ (and also from parameters $\theta_i$ and $\phi_i$) because only one subsystem is considered. It is interesting to notice that, it is possible to have an input-independent parameter. This can be done by computing the average linear entropy with respect to all possible input states on the surface of the Bloch sphere as:
\begin{equation}
S_{_A}^{\text{av}}(t)=\int S_A(\theta,\phi;t) \, \text{d}\Omega,
\end{equation}
in which $\text{d}\Omega$ is the normalized $SU(2)$ Haar measure. This is related to the concept of entangling power \cite{Zanardi2000}. 

According to (\ref{eq:survivalamplitude}) and (\ref{eq:red}), the effective dynamics of the linear entropy depends on the function ${\cal E}(t)$. It is worth noticing that, from  Eq. (\ref{eq:corrfunc}) two distinct weak and strong coupling regimes can be distinguished by introducing the dimensionless parameter  $R=g/\kappa$, by which we are able to analyse our results in two regimes, good ($R\gg 1$) and bad ($R\ll 1$) cavities. In the bad cavity limit, the relaxation time is greater than the reservoir correlation time and the variation of linear entropy is essentially a Markovian exponential behaviour.
In the good cavity limit, the reservoir correlation time is greater than the relaxation time and non-Markovian effects such as revival and oscillation of entanglement become dominant. These latter effects are due to the long memory of the environment.

\begin{figure}[h!]
\centering
\subfigure[\label{Fig_2a} ]{\includegraphics[width=0.35\textwidth]{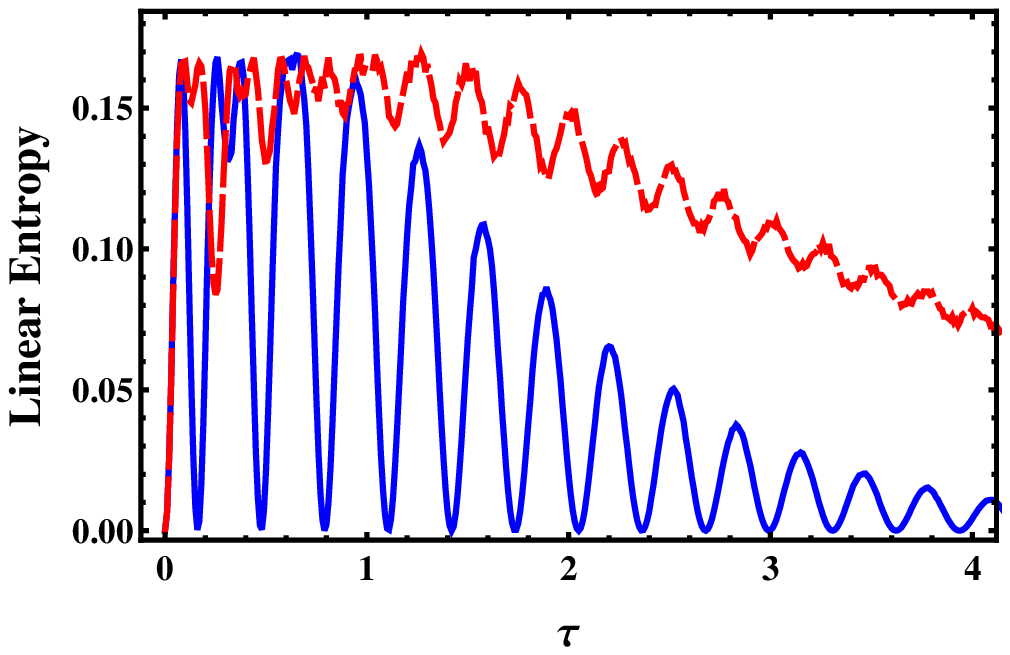}}
\hspace{0.05\textwidth}
\subfigure[\label{Fig_2b} ]{\includegraphics[width=0.35\textwidth]{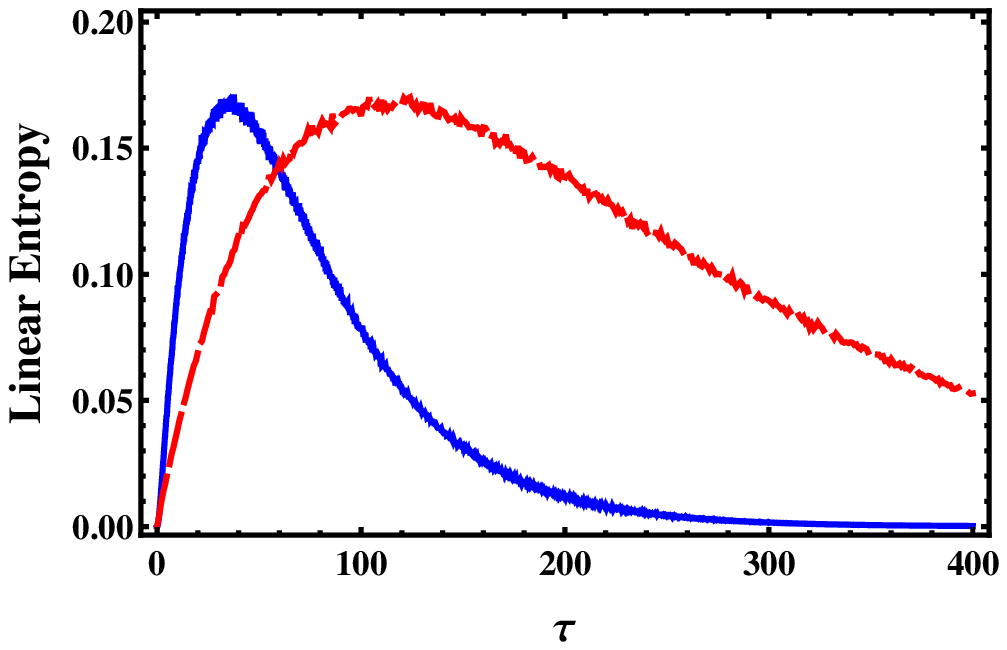}}

\caption{Time evolution of the average of linear entropy as function of scaled time $\tau$ for (a) strong coupling regime, i.e., $R=10$ with $\Delta=0$ (solid blue line) and $\Delta=15\kappa$ (dashed red line) and (b) weak coupling regime, i.e., $R=0.1$ with $\Delta=0$ (solid blue line) and $\Delta=1.5\kappa$ (dashed red line).} \label{Fig_2}
   \end{figure}

Figure \ref{Fig_2} illustrates the average linear entropy as function of the scaled time $\tau=\kappa t$ for both strong and weak coupling regimes in the absence and presence of detuning. In the strong coupling regime and in the absence of the detuning parameter, the entropy has an oscillatory decaying behaviour which represents a non-Markovian process. 
These revivals and oscillations are due to the memory depth of the reservoir. This can be understood from the fact that, in the strong coupling regime, the environment feedbacks part of the information which it has taken during the interaction with the qubit. In the presence of detuning parameter, the entanglement sudden death is no longer seen. The linear purity remains alive at longer intervals of time. On the other hand, in the weak coupling regime, linear entropy starts from zero and increases monotonically up to its maximum value and then it falls down and decreases until it vanishes. Again, the detuning parameter makes the entropy to be survive in longer times. Actually, for both coupling regimes and for sufficient high values of detuning parameter, it is possible to have a quasi-stationary entanglement. 

\section{Entanglement Swapping}\label{sec.Etp}

As is observed, there is no direct interaction among the two  $(\mathrm{AF})_i$ systems, therefore their states are expected to remain separable:
         \begin{equation}\label{statet}
         \hat{\rho}(t)= \ket{\Psi(t)}\bra{\Psi(t)},
       \end{equation}
 in which $\ket{\Psi(t)}=\ket{\psi_{\mathrm{AF}}(t)}_1\otimes\ket{\psi_{\mathrm{AF}}(t)}_2$. However, thanks to the results of the previous section, where it is established that the states of atom-field are entangled  in each cavity. Now, in the line of the goals of our paper, it is quite reasonable to search for a strategy to exchange the entanglement between atom-field in each cavity into atom-atom (and/or field-field) for the sake of quantum information processing purposes. In this regard, one could think of creating entanglement between the two atoms by performing the BSM onto the field modes leaving the cavities (see Fig. \ref{Fig_1}). Mathematically speaking, it can be done by projection $\ket{\Psi(t)}$ onto one of the Bell states of the cavity fields. Among different types of resources for linear optical quantum swapping implementations, the two-photon pairs has been put forward to be an efficient resource for this purpose which are \cite{Lee2013}:
\begin{subequations}
\begin{eqnarray}
 \ket{\Psi^{\pm}}_\mathrm{F}&=&\frac{1}{\sqrt{2}}\left( \ket{\boldsymbol{0}}_{R_1}\ket{\boldsymbol{1}}_{R_2}\pm\ket{\boldsymbol{0}}_{R_1}\ket{\boldsymbol{1}}_{R_2} \right), \\
 \ket{\Phi^{\pm}}_\mathrm{F}&=&\frac{1}{\sqrt{2}}\left( \ket{\boldsymbol{0}}_{R_1}\ket{\boldsymbol{0}}_{R_2}\pm\ket{\boldsymbol{1}}_{R_1}\ket{\boldsymbol{1}}_{R_2} \right), 
\end{eqnarray}
\end{subequations}
in which $\ket{\boldsymbol{0}}_{R_i}$ has been defined before and 
\begin{equation}
\ket{\boldsymbol{1}}_{R_i}\equiv\int\! \Theta(\omega)\ket{1_{\omega}}_i \ \text{d}\omega
\end{equation}
where $\int\! |\Theta(\omega)|^2 \text{d}\omega=1$ with $\Theta(\omega)$ as the pulse shape associated with the incoming photon. Using the introduced Bell-type states, one can easily construct the desired projection operator $P_\mathrm{F}=\ket{M}{}_\mathrm{FF}\bra{{}M}$ in which $M\in\left\lbrace \Psi^{\pm},\Phi^{\pm}\right\rbrace$. Consequently,  operating the projection operator on $\ket{\Psi(t)}$ leaves the field states in a Bell-type state and also establish entangled atom-atom state. In the next two subsections, we shall consider the projection operator based on Bell states $\ket{\Psi^-}$ and $\ket{\Phi^+}$ and investigate the resulting entanglement properties of the atom-atom states. It should be noted that the other two Bell states can also straightforwardly be taken into account.

\subsection{Bell state $\ket{\Psi^{-}}_\mathrm{F}$}

Let us now consider the following projection operator 
\begin{equation}
P^-_\mathrm{F}=\ket{\Psi^-}{}_\mathrm{FF}\bra{{}\Psi^-},
\end{equation}
which its action on $\ket{\Psi(t)}$ in (\ref{statet}) leaves the field states in the Bell state $\ket{\Psi^-}{}_\mathrm{F}$ and establishes the following atom-atom state (after normalization):
\begin{equation}\label{AAunnormPsi}
\begin{aligned}
     \ket{\Psi_\mathrm{AA}(t)}&=P^-_F \ket{\Psi(t)}  \\
     &=\frac{1}{\sqrt{N^-(t)}}\left\lbrace  X_{12}(t)\ket{e,g}-X_{21}(t)\ket{g,e}\right. \\
     &+\left. \left( \Upsilon_{12}(t)-\Upsilon_{21}(t)\right) \ket{g,g}\right\rbrace,
\end{aligned}
\end{equation}
in which the normalization coefficient reads as
\begin{equation}\label{NorCoeffPsi}
\begin{aligned}
     N^-(t)&= |X_{12}(t)|^2+|X_{21}(t)|^2+|\Upsilon_{12}(t)-\Upsilon_{21}(t)|^2.
\end{aligned}
\end{equation}
Here, we have defined
\begin{subequations}
\begin{eqnarray}
X_{jk}(t)&=& C_j(t)\int\! \textbf{d}\omega\Theta^{*}(\omega)U_{\omega_{_k}}(t)e^{-i\omega t}, \label{eq:Theta} \\
\Upsilon_{jk}(t)&=& D_j(t)\int\! \textbf{d}\omega\Theta^{*}(\omega)U_{\omega_{_k}}(t)e^{-i\omega t}. \label{eq:Upsilon}
\end{eqnarray}
\end{subequations}
In order to quantify the amount of entanglement between the two atoms, we use the concurrence \cite{Wootters1998}  which has been defined as
       \begin{equation}\label{con}
        E\left( \hat{\rho}(t)\right) =\mathrm{max}\{0,\sqrt{\lambda_1}-\sqrt{\lambda_2}-\sqrt{\lambda_3}-\sqrt{\lambda_4}\},
       \end{equation}
where $\lambda_i$, $i=1,2,3,4$ are the eigenvalues (in decreasing order) of the Hermitian matrix
$\hat{\rho}_{_\mathrm{AA}}\left(\sigma_1^y\otimes\sigma_2^y\hat{\rho}_{_\mathrm{AA}}^{*}\sigma_1^y\otimes\sigma_2^y\right)$ with $\hat{\rho}_{_\mathrm{AA}}^*$ the complex conjugate of $\hat{\rho}_{_\mathrm{AA}}$ and $\sigma_k^y:=i(\sigma_k-\sigma_k^\dag)$.  The concurrence varies between 0 (completely separable) and 1 (maximally entangled).
For state (\ref{AAunnormPsi}), the concurrence reads as
\begin{equation}
E\left( \hat{\rho}(t)\right)=\dfrac{T_1(t,\theta_1,\theta_2)}{T_1(t,\theta_1,\theta_2)+T_2(\theta_1,\theta_2,\phi_1,\phi_2)},
\label{conexPsi}
\end{equation}
in which, 
\begin{subequations}
\begin{eqnarray}
T_1(t,\theta_1,\theta_2)&=& 2\cos^2(\theta_1/2)\cos^2(\theta_2/2)\left|  {\cal E}(t)\right|  ^2, \label{eq:T1} \\
T_2(\theta_1,\theta_2,\phi_1,\phi_2)&=& \dfrac{1}{2}\Big( 1-\cos\theta_1\cos\theta_2 \nonumber \\
&-&\sin\theta_1\sin\theta_2\cos(\phi_1-\phi_2)\Big). \label{eq:T2}   
\end{eqnarray}
\end{subequations}

\begin{figure}[h!]
\centering
\subfigure[\label{Fig_3a} ]{\includegraphics[width=0.35\textwidth]{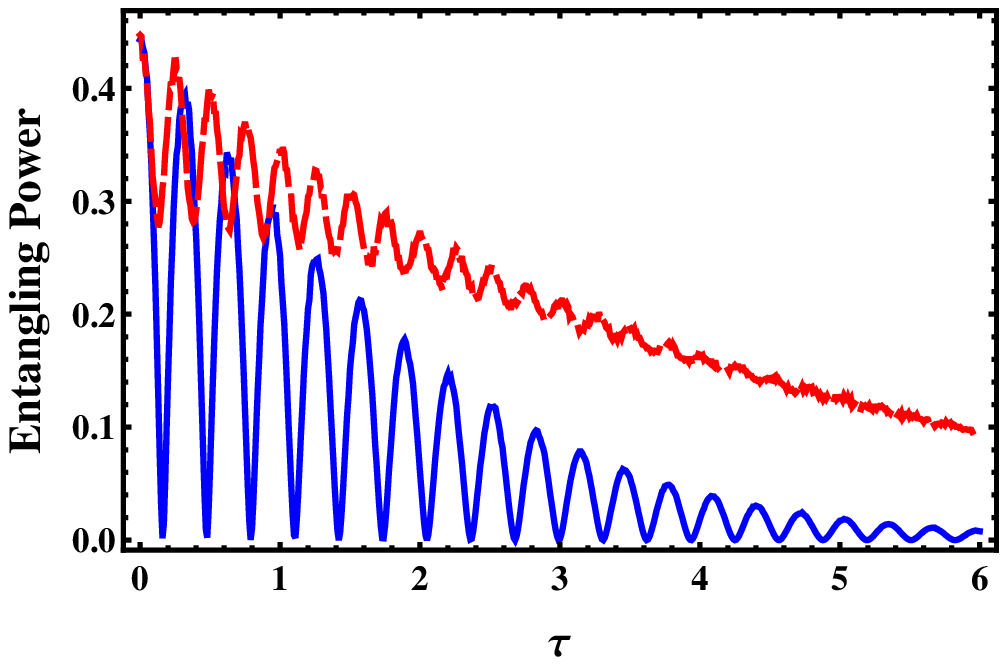}}
\hspace{0.05\textwidth}
\subfigure[\label{Fig_3b} ]{\includegraphics[width=0.35\textwidth]{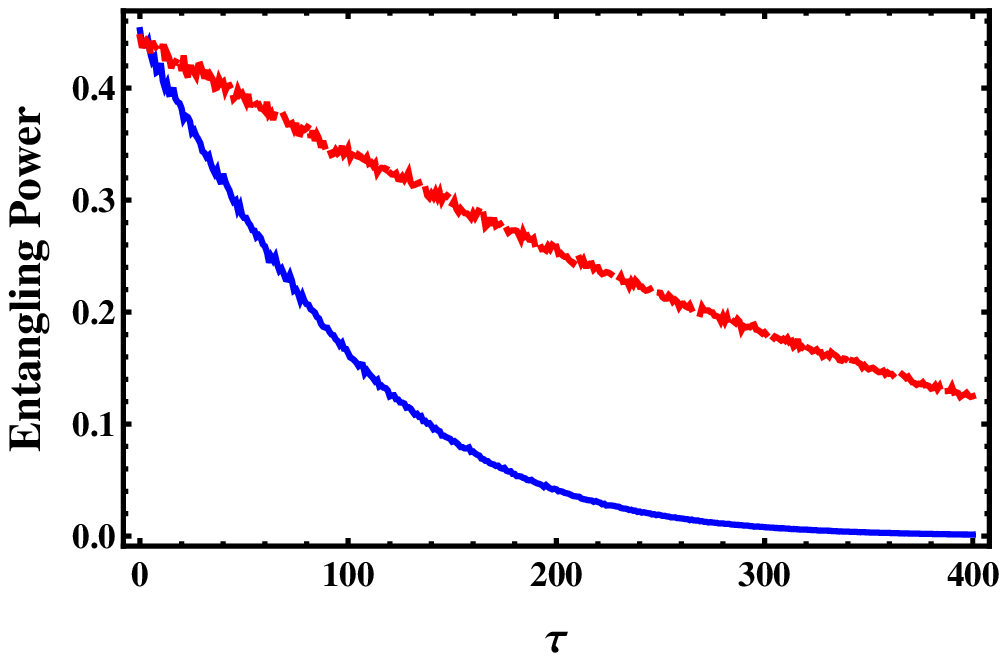}}
\caption{Time evolution of the entangling power of the atom-atom state after BSM ($P^-_\mathrm{F}=\ket{\Psi^-}{}_\mathrm{FF}\bra{{}\Psi^-}$) as function of scaled time $\tau$ for (a) strong coupling regime, i.e., $R=10$ with $\Delta=0$ (solid blue line) and $\Delta=15\kappa$ (dashed red line) and (b) weak coupling regime, i.e., $R=0.1$ with $\Delta=0$ (solid blue line) and $\Delta=1.5\kappa$ (dashed red line).}\label{Fig_3}
   \end{figure}

The surprising aspect here is that the resulting concurrence does not depend on the pulse shape of the incoming photon (i.e., $\Theta(\omega)$). Before considering the time evolution of the resulting concurrence, it is interesting to notice that, for certain conditions, the state (\ref{AAunnormPsi}) can have a unique stationary state. A glance at (\ref{conexPsi}) reveals that whenever $T_2(\theta_1,\theta_2,\phi_1,\phi_2)=0$, the concurrence would be independent of time and it remains always at its maximum value, i.e., 1. According to (\ref{eq:T2}), with the following set of solutions this condition is fulfilled:
\begin{equation}
\theta_1 =\theta_2 \ \text{and} \ \phi_1-\phi_2=2m\pi, \ \ m=0,\pm 1,
\end{equation}
which leads to the maximally entangled Bell state (up to an irrelevant global phase)
  \begin{eqnarray}\label{Bellstate1}
  \ket{\Psi^-}=\frac{1}{\sqrt{2}}(\ket{e,g}-\ket{g,e}).
  \end{eqnarray}  
Quite generally, the concurrence (\ref{conexPsi}) depends on the initial state.  However, it is logical to state that our model is a good entangler when the average of the final swapped entanglement over all possible initial states is positive. This statistical average over the initial states establishes an input-independent dynamics of entanglement. Therefore, we use the concept of entangling power which is defined as \cite{Zanardi2000}
\begin{equation}
{\mathfrak E}(t):=\int E\left(\rho(t)\right) \, d\mu( |\psi(0)\rangle),
\label{eq:enpower}
\end{equation}
where $ d\mu( |\psi(0)\rangle)$ is the probability measure over the submanifold of product states in $\mathbb{C}^2\otimes \mathbb{C}^2$. The latter is induced by the Haar measure of ${\rm SU}(2) \otimes {\rm SU}(2)$. Specifically, referring to the parametrization of \eqref{eq:initialstate}, it reads
\begin{equation}
d\mu( |\psi(0)\rangle)=\frac{1}{16\pi^2}\prod\limits_{k=1}^2 \sin\theta_k\text{d}\theta_k\text{d}\varphi_k.
\end{equation}
This measure is normalized to 1. It is trivial to see that in this case the entangling power $\mathfrak E$ lies in $[0,1]$.\\
As is stated before, two distinct strong and weak coupling regimes can be distinguished.
Fig. \ref{Fig_3} illustrates the entangling power as function of scaled time $\tau=\kappa t$ in the absence and presence of detuning for both coupling regimes.
In the strong coupling regime, an oscillatory behaviour of entanglement is seen due to the long memory effect of the cavities. In both cases the entangling power has a decaying behaviour in the absence and presence of detuning parameter. The detuning parameter has a crucial role in surviving the swapped entanglement. Specially, in the strong coupling regime, it completely suppresses the entanglement sudden death.

\subsection{Bell state $\ket{\Phi^{+}}_\mathrm{F}$}
\begin{figure}[h!]
\centering
\subfigure[\label{Fig_4a}]{\includegraphics[width=0.35\textwidth]{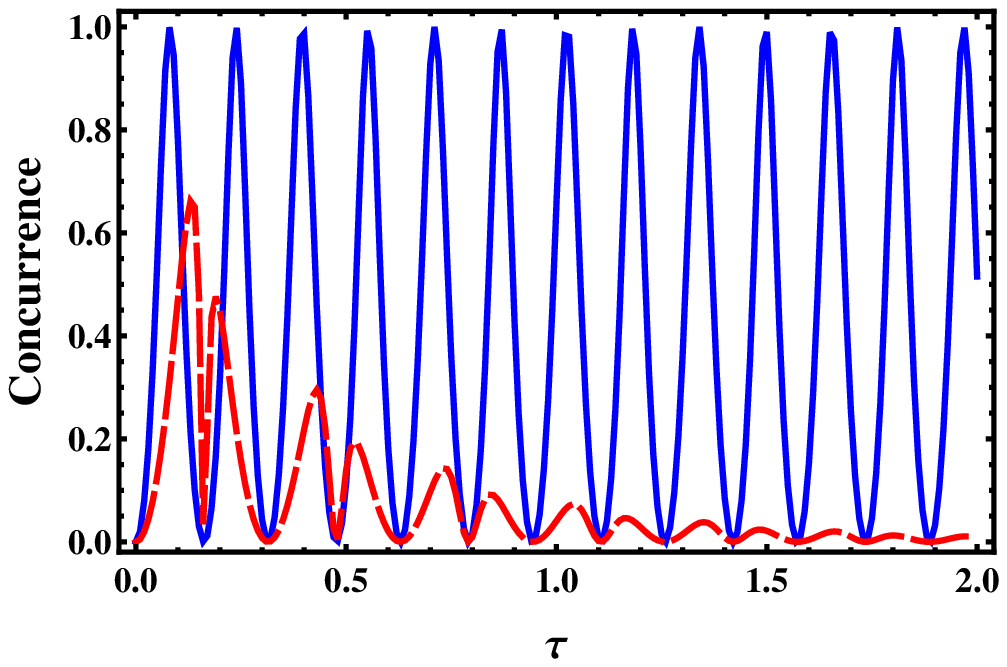}}
\hspace{0.05\textwidth}
\subfigure[\label{Fig_4b}]{\includegraphics[width=0.35\textwidth]{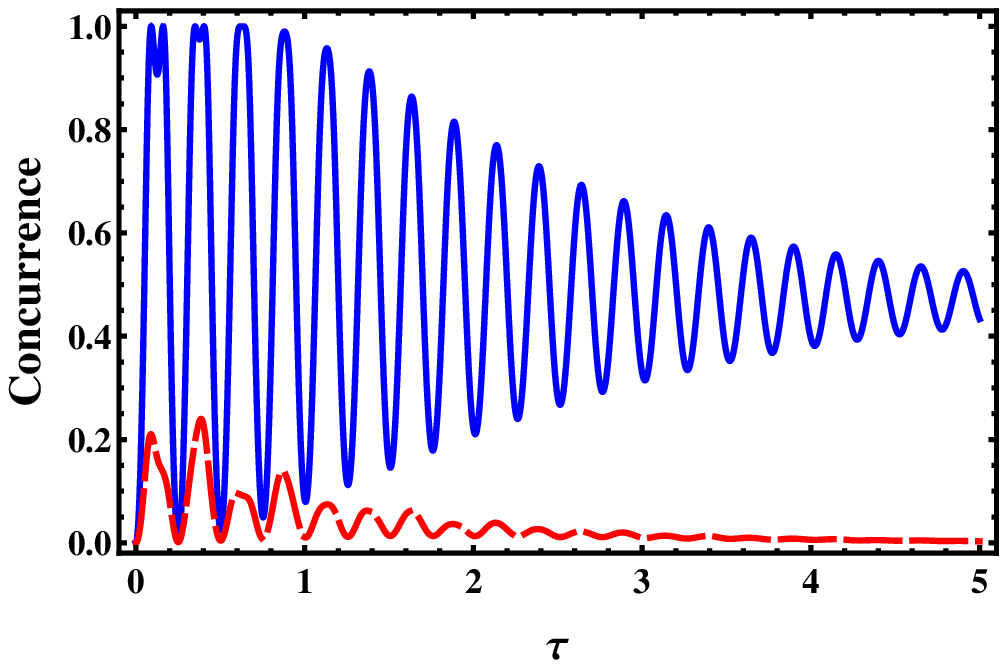}}
\caption{Time evolution of the concurrence of the atom-atom state after BSM ($P^+_\mathrm{F}=\ket{\Phi^+}{}_\mathrm{FF}\bra{{}\Phi^+}$) as function of scaled time $\tau$ for strong coupling regime, i.e., $R=10$ for (a) $\Delta=0$ and (b) $\Delta=15\kappa$ with $\theta_1=\theta_2=0$ (blue solid line) and $\theta_1=\theta_2=\pi/2$ and $\phi_1=\phi_2=0$ (red dashed line).}\label{Fig_4}
   \end{figure}

Let us now consider another Bell state, i.e., $\ket{\Phi^+}_\text{F}$ in order to construct projection operator $P^+_\mathrm{F}=\ket{\Phi^+}{}_\mathrm{FF}\bra{{}\Phi^+}$. It is straightforward to obtain the following (normalized) atom-atom state by  acting this new projection operator onto state (\ref{statet})
\begin{equation}\label{AAunnorm}
\begin{aligned}
     \ket{\Psi_\mathrm{AA}(t)}&= P^+_F\ket{\Psi(t)}  \\
     &=\frac{1}{\sqrt{N^+}}\Big(  C_1(t)C_2(t)\ket{e,e}+C_1(t)D_2(t)\ket{e,g} \\
     &+ D_1(t)C_2(t)\ket{g,e} \\
     &+ \left( D_1(t)D_2(t)+C_1(0)C_2(0)\Gamma(t)^2\right)  \ket{g,g} \Big),
\end{aligned}
\end{equation}
in which
\begin{equation}\label{Gamma}
\Gamma(t)=-ig\int_{0}^{t}\!\mathrm{d}t_1{\cal E}(t_1)e^{i\omega_{\text{qb}}(t-t_1)}\int\!\text{d}\omega\alpha(\omega)\Theta^{*}(\omega)e^{-i\omega(t-t_1)}. 
\end{equation}
In the above relations, the normalization coefficient is
\begin{equation}\label{NorCoeff}
\begin{aligned}
      N^+&\equiv N^+(t,\theta_1,\theta_2,\phi_1,\phi_2) \\
      &=\cos^2(\theta_1/2)\cos^2(\theta_2/2)\left( \left|  {\cal E}(t)\right|  ^4+\left|  \Gamma(t)\right|  ^4\right) \\
      &+0.5\left|  {\cal E}(t)\right|  ^2\left( 1-\cos\theta_1\cos\theta_2\right) +\sin^2(\theta_1/2)\sin^2(\theta_2/2)\\
      &-0.5\sin\theta_1\sin\theta_2\Re\left( e^{-i(\phi_1+\phi_2)}\Gamma^2(t)\right). 
\end{aligned}
\end{equation}
For state (\ref{AAunnorm}) the resulting concurrence explicitly reads as
\begin{equation}
E\left( \hat{\rho}(t)\right)=\dfrac{T(t,\theta_1,\theta_2)}{N^+(t,\theta_1,\theta_2,\phi_1,\phi_2)},
\label{conex}
\end{equation}
in which
\begin{equation}
T(t,\theta_1,\theta_2)= 2\cos^2(\theta_1/2)\cos^2(\theta_2/2)\left|  {\cal E}(t)\right|  ^2\left|  \Gamma(t)\right|  ^2. \label{eq:T}
\end{equation} 
Unlike the previous case, the resulting concurrence depends on the pulse shape of the incoming photon, i.e., $\Theta(\omega)$. Therefore, different pulse shapes lead to different behaviour of the dynamics of entanglement. However, due to the technical difficulties arising when calculating the integrals (\ref{Gamma}), we consider the incoming pulse shape exactly the same as (\ref{eq:alpha}). With this assumption, the function $\Gamma(t)$ explicitly becomes
\begin{equation}\label{GammaLo}
\Gamma(t)=-2ie^{-(i\Delta+\kappa)t/2}\dfrac{g}{\Omega}\sinh(\Omega t/2).
\end{equation} 
Quite generally, the concurrence (\ref{conex}) is zero at any time for $\theta_1=\theta_2=\pi$. This condition corresponds to  atomic initial state $\ket{g,g}$ (up to a global phase). Here, no entanglement can be created because no excitation can be exchanged between two qubits
by the action of BSM. Moreover, according to (\ref{AAunnorm}) and (\ref{NorCoeff}) the final state after BSM is $\ket{g,g}$ (up to an irrelevant global phase) which clearly is a separable state. \\
On the other hand, we expect that the concurrence reaches its maximum value for $\theta_1=\theta_2=0$. This corresponds to the initial atomic state $\ket{e,e}$. It is straightforward to show that with these values, the concurrence is maximum whenever the following condition is fulfilled
\begin{equation}\label{condition}
\left| {\cal E}(t)\right| =\left| \Gamma(t)\right|.
\end{equation}
However, due to the presence of parameter $g$ in the expression of $\Gamma(t)$, this condition can only be fulfilled in the strong coupling regime. In order to see this phenomenon explicitly, let us plot the concurrence for some initial states in both coupling regimes.   \\
In Fig. \ref{Fig_4} we have plotted the time evolution of the concurrence (\ref{conex}) as function of the scaled time $\tau=\kappa t$ in the strong coupling regime (i.e., $R=10$) for two atomic initial states in the absence and presence of detuning parameter. Let us first investigate the case in which the exact resonance condition is considered. As is seen from Fig. \ref{Fig_4a} (solid plot), for initial atomic state $\ket{e,e}$ (i.e., $\theta_1=\theta_2=0$) and in the absence of detuning parameter ($\Delta=0$), the concurrence has an oscillatory behaviour between zero and its maximum value (i.e., 1). Therefore, despite of the presence of dissipation, it is possible to achieve the maximum amount of entanglement. The maximum value of entanglement is obtained whenever the condition (\ref{condition}) is fulfilled which straightforwardly arrives us at the following (scaled) times $\tau_n$ at which the concurrence is maximum:
 \begin{equation}\label{Tnteta0}
 \tau_n=\frac{1}{10}\left( 2n\pi+\dfrac{\pi}{4} \right),
 \end{equation}
where $n$ is an integer. At these times, it is straightforward to show that, ${\cal E}(t)$ and $\Gamma(t)$ become real functions of time and consequently the atom-atom state is projected into the following Bell state:
\begin{eqnarray}\label{Bellstate}
 \ket{\Psi_\mathrm{AA}}=\frac{1}{\sqrt{2}}(\ket{e,e}+\ket{g,g}).
\end{eqnarray}
For another atomic initial state (i.e., $\theta_1=\theta_2=\pi/2$) and in the absence of detuning parameter, an oscillatory and decaying behaviour of concurrence is seen (see Fig. \ref{Fig_4a} dashed plot). The entanglement sudden death phenomenon is clearly seen and no stationary entanglement is created.  \\
In the presence of the detuning parameter and for $\theta_1=\theta_2=0$, the oscillatory and decaying behaviour of the concurrence is clearly seen. However, the entanglement sudden death is no longer observed. As the scaled time goes on, the amplitude of oscillations decreases until the concurrence reaches the stationary value (here it is $0.47$). Our further calculations (not shown here) illustrate that, as the detuning parameter increases, the stationary value of concurrence decreases. For other initial state, the detuning parameter makes concurrence to vanish in shorter scaled times. 
   
\begin{figure}[ht]
   \centering
\includegraphics[width=0.35\textwidth]{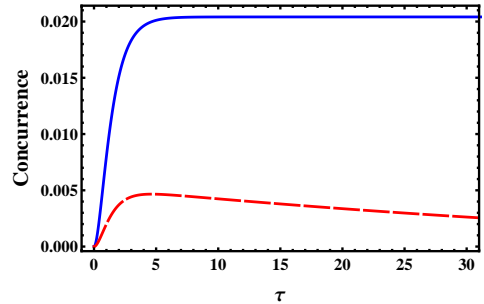}
   \caption{\label{Fig_5} Time evolution of the concurrence of the atom-atom state after BSM ($P^+_\mathrm{F}=\ket{\Phi^+}{}_\mathrm{FF}\bra{{}\Phi^+}$) as function of scaled time $\tau$ for weak coupling regime, i.e., $R=0.1$ in the absence of detuning parameter (i.e., $\Delta=0$) for $\theta_1=\theta_2=0$ (blue solid line) and $\theta_1=\theta_2=\pi/2$ and $\phi_1=\phi_2=0$ (red dashed line).}
  \end{figure}

\begin{figure}[h!]
\centering
\subfigure[\label{Fig_6a} ]{\includegraphics[width=0.35\textwidth]{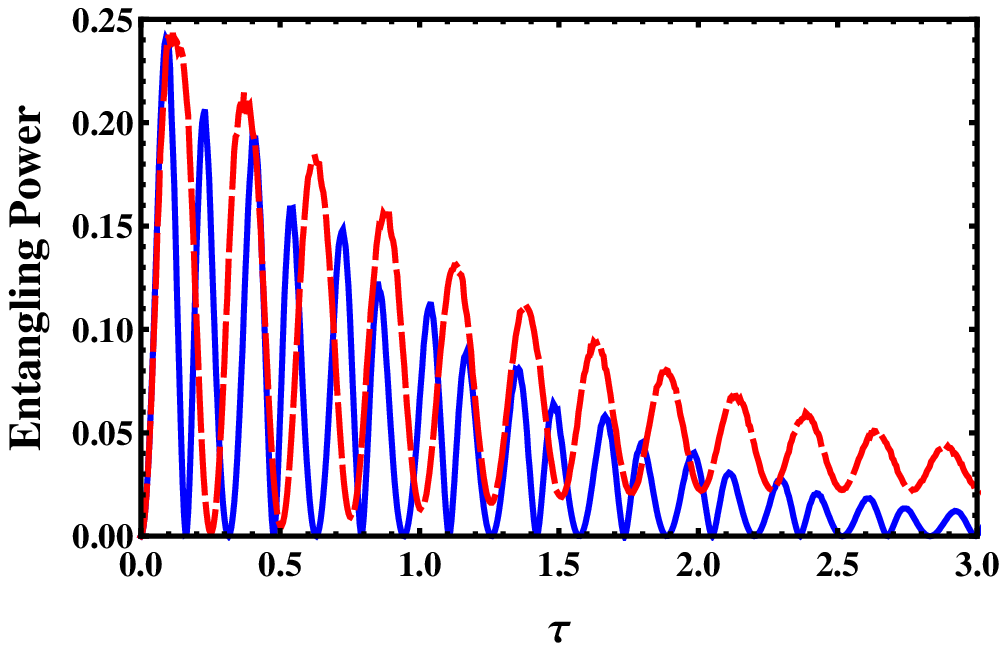}}
\hspace{0.05\textwidth}
\subfigure[\label{Fig_6b} ]{\includegraphics[width=0.35\textwidth]{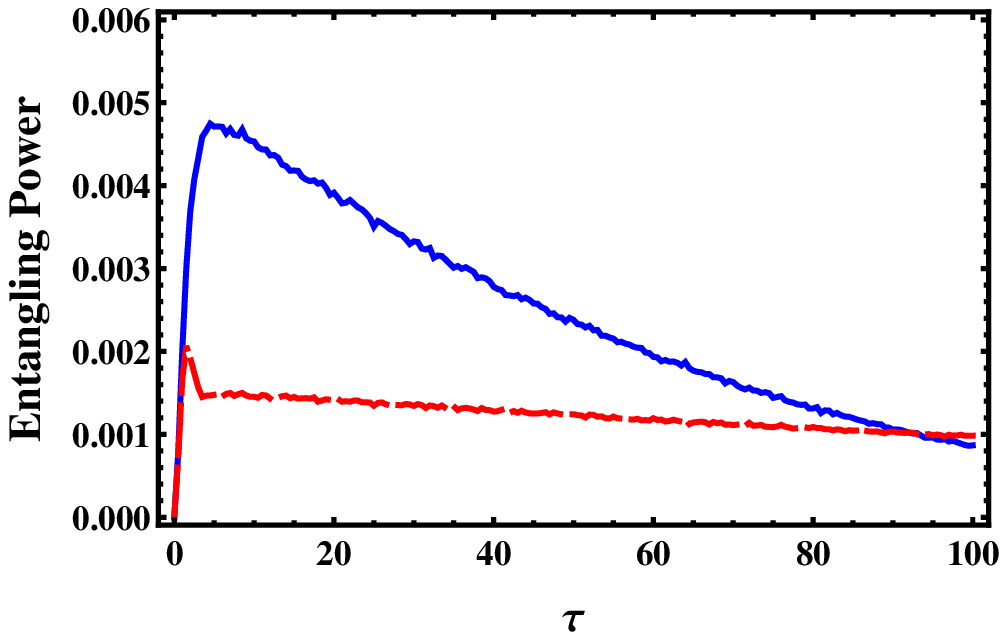}}
\caption{Time evolution of the entangling power of the atom-atom state after BSM as function of scaled time $\tau$ for (a) strong coupling regime, i.e., $R=10$ with $\Delta=0$ (solid blue line) and $\Delta=15\kappa$ (dashed red line) and (b) weak coupling regime, i.e., $R=0.1$ with $\Delta=0$ (solid blue line) and $\Delta=1.5\kappa$ (dashed red line).}\label{Fig_6}
   \end{figure}

Figure \ref{Fig_5} illustrates the resulting concurrence as function of the scaled time $\tau$ in weak coupling regime for zero detuning parameter with different atomic initial states. 
The weak coupling regime shows different behaviour. In this regime and for $\theta_1=\theta_2=0$, the concurrence starts from zero and  increases monotonically up to the stationary value and then remains at this value as time goes on. The amount of the swapped entanglement is negligible in comparison with the strong coupling regime. This can be explained by paying attention to the fact that in the weak coupling regime, the correlation between a typical qubit and its environment is too weak. Therefore, the amount of swapped entanglement must be less than the one in the strong coupling regime. As explained before, the concurrence never reaches its maximum value in this regime. For another initial state, the stationary concurrence no longer exists. Finally, it should be noted that our further calculations  show that the presence of  detuning parameter decreases considerably the amount of the stationary concurrence. \\
After having ascribed the role of initial state on the dynamics of swapped entanglement, we intend to examine, on average, how much entanglement can be swapped between two qubits. For this purpose, we have plotted the entangling power as a function of the scaled time $\tau=\kappa t$ for two strong and weak coupling regimes in the absence and presence of detuning parameter (see Fig. \ref{Fig_6}). In the strong coupling regime, an oscillatory behaviour of entangling power with a decaying envelop is clearly observed. In the absence of the detuning parameter the entanglement sudden death phenomenon is clearly seen. \\
In the weak coupling regime, on average, the amount of  entanglement is negligible in comparison with the strong coupling regime. However, a small amount of entanglement is seen in the presence of detuning parameter. 
 
\section{Concluding Remarks}\label{sec.Con}

To sum up, we considered two independent dissipative cavities, each consists of a qubit. In each dissipative cavity, the qubit interacts with a single-mode field and the field interacts with a set of continuum  harmonic oscillators. Therefore, the leakage of photons into a continuum of states is the source of dissipation. This allows us to investigate our results outside of Markovian limit. However, by introducing new set of dressed operators $\hat{A}_i(\omega)$ which contain the information of the cavity field and the surrounding environment, we solved the time-dependent Schrödinger equation and obtained the analytical expression of the wave vector of each subsystem for special initial conditions.

Then, before considering the entanglement swapping protocol, we investigated the dynamics of entanglement between each qubit and its surrounding environment. This has been done by evaluating the average of linear entropy measure over all possible initial states of the qubit. The results show that an oscillatory decaying behaviour of entropy is seen in the strong coupling regime and in the absence of detuning parameter. These oscillations are due to the long memory of the environment. However, this behaviour is not occurred in the weak coupling regime, where the behaviour of entropy is a monotonically decaying one. In the presence of detuning parameter, the linear entropy survives at longer intervals of time for both regimes.

After that, we have implemented the entanglement swapping protocol to transform entanglement from two
atom-field subsystems to atom-atom by an interference measurement performed on the fields leaving the cavities (BSM). This has been done by projecting the state of entire system onto one of the Bell states of the cavity fields. We have presented our results for two types of the field-field Bell-type states. First, we considered $\ket{\Psi^{-}}_\mathrm{F}$ Bell state. In this case, the resulting concurrence does not depend on the pulse shape of incoming photons. Furthermore, for $\theta_1 =\theta_2 \ \text{and} \ \phi_1-\phi_2=2m\pi \ (m=0,\pm 1)$, the atom-atom state have the unique stationary state $  \ket{\Psi^-}=\frac{1}{\sqrt{2}}(\ket{e,g}-\ket{g,e}).$ In order to investigate the dynamical behaviour of swapped entanglement, we introduced the entangling power (\ref{eq:enpower}). Again, an oscillatory behaviour of entanglement is seen for strong coupling regime. In both regimes, the swapped entanglement has a decaying behaviour. However, the detuning parameter play a crucial role in surviving the swapped entanglement.
 
On the other hand, for field-filed Bell state $\ket{\Phi^{+}}_\mathrm{F}$, the situation  differs slightly. First of all, the resulting concurrence depends directly on the pulse shape of incoming photons. By assuming that the pulse shape is the same as $\alpha(\omega)$, it is possible to solve the relevant integrals and obtain the analytical expression for concurrence. Second, unlike the previous case, there is not a unique entangled stationary state, but for $\theta_1=\theta_2=\pi$ the concurrence is zero at any time $t$.  
In the strong coupling regime and for initial atomic state $\ket{e,e}$ (i.e., $\theta_1=\theta_2=0$) and in the absence of detuning parameter ($\Delta=0$),  an oscillatory (without decaying) behaviour of concurrence is seen. Our results show that at discrete (scaled) times $ \tau_n=\frac{1}{10}\left( 2n\pi+\dfrac{\pi}{4} \right)$ where $n$ is an integer, the concurrence reaches its maximum value and the atom-atom state is projected into the maximally entangled Bell state $\frac{1}{\sqrt{2}}(\ket{e,e}+\ket{g,g})$. Furthermore, the amount of swapped entanglement in the weak coupling regime is negligible in comparison with the strong coupling regime.

Finally, we should state that our results could be helpful in designing experiments for entanglement swapping when the environmental effects cannot be neglected. For example, each qubit-field subsystem in a dissipative cavity can be considered as polarization of a decohered photon. Then, two polarization-entangled photon pairs can be generated by spontaneous parametric down-conversion \cite{Yamamoto2003}.
Furthermore, entanglement swapping plays a crucial rule in development of real devices for applications of quantum information theory, such as development of quantum computers where subsystems unavoidably interact with their environment. Therefore, our analytical results are expected to be a first step towards that goals.

 \section*{References}
  \bibliographystyle{apsrev4-1}
  \bibliography{PRARefBank}                 
\end{document}